\documentclass[journal]{IEEEtran}

\usepackage{graphicx}
\usepackage[justification=centering]{caption}
\usepackage{caption}
\usepackage{cite}
\usepackage{booktabs}

\begin{document}

\title{Small World Model for scaling up prediction result based on SEIR model}

\author{Guixu Lin, Defan Feng, Peiran Li, Yicheng Zhao, Haoran Zhang, Xuan Song\textsuperscript{*} 
\thanks{*Corresponding author at: SUSTech-UTokyo Joint Research Center on Super Smart City, e-mail addresses: songx@sustech.edu.cn (X. Song)}

\thanks{
Guixu Lin, Yicheng Zhao and Xuan Song are with SUSTech-UTokyo Joint Research Center on Super Smart City, Southern University of Science and Technology}
\thanks{Defan Feng is with the Department of Computer Science and Engineering, Southern University of Science and Technology}

\thanks{Guixu Lin, Peiran Li, Haoran Zhang are with Center for Spatial Information Science, University of Tokyo}
}

\maketitle

\begin{abstract}
Data-driven epidemic simulation helps better policymaking. Compared with macro-scale simulations driven by statistical data, individual-level GPS data can afford finer and spatialized results. However, due to the difficulty of data collection and acquisition, GPS data used in epidemiological studies cannot cover all populations, reducing the credibility of epidemiological simulation results. Therefore, this paper proposes a Small World Model to more accurately map the results from the "small world" (simulation with partially sampled data) to the real world. Based on the basic principles of disease transmission, this study derives two parameters: time scaling factor and number scaling factor. Time scaling factor indicates the 
spread rate in the epidemic transmission model, while the number scaling factor maps the simulated infected number to the real infected number. The small World Model not only converts the simulation of the small world into the state of the real world but also analyzes the effectiveness of different mobility restriction policies.
\end{abstract}




\IEEEpeerreviewmaketitle
\section{Introduction}
\subsection{Background}
The epidemic transmission is highly correlated with human mobility \cite{kraemer2019utilizing,charaudeau2014commuter,mari2012role}. In the early stage of an epidemic outbreak, adopting effective human mobility restriction policies, such as curfew or teleworking, can effectively curb the spread of virus. However, it is very hard for government to adopt novel policies. The difficulty lies in a lack of understanding of aforesaid disease and no globally acknowledged metrics to evaluate restriction policies. In general, building an epidemic transmission simulation model can predict infected under different policies and support government to formulate effective and manageable policies. Such simulation models are usually driven by human mobility data as well as confirmed cases data. Models utilizing individual level data, such as individual trajectory data, have a higher resolution and accuracy than models using statistic data. Based on the simulation, urban policy makers can assign different prevention policies for different regions with the help of simulation result. 

Currently, the study of epidemic transmission model driven by individual level data has a major challenge: the reliability of result.
Reliability depends on the scale of data. If a dataset contains mobility of the whole population, the simulation of the epidemic transmission model can completely represent the status of real world. However, in most cases, it is impossible to collect and represent the mobility of all population. Researchers take the simulation based on a partial, limited data. This kind of partial data is called small dataset in the paper, while epidemic simulation model based on partial data is called small world model. The data limitation in small world model bias the credibility of the simulation results. Although simulations of small world model cannot directly represent that of real world, there is a correlation between the small world and the real world. 
Rather than simply multiplying the result of small world model by the population ratio 
between small world and real world, how to more accurately map the spread in small world model to real world is important because it would improve the simulation accuracy based on small dataset.

To address the aforementioned challenge,  based on individual level data, the paper proposes the method named Small World Model to convert the simulation result of small world into that of real world.

\subsection{Literature Review}
Many studies have established data driven transmission models to simulate the spread of infectious diseases, which are used to analyze the effectiveness of travel restrictions and other policies. According to the granularity of data used, it can be roughly divided into two types of research.

The first type uses statistical level data to simulate the epidemic transmission. Chinazzi et al. \cite{chinazzi2020effect} used two kinds of statistical level mobility data including the airline transportation data and population flow data to build a model named GLEAM to analyze the effect of travel restrictions on epidemical transmission. The work of \cite{kucharski2020early,fang2020transmission,liu2020modelling} used SEIR transmission model based on the reported case data of COVID-19 to analyze the effectiveness of government interventions during COVID-19. Utilizing the daily COVID-19 incidence data and airline data, Wells et al. \cite{wells2020impact} used Monte Carlo simulations to estimate the effect of border control measures of COVID-19 outbreak. All these works provide the global tendency of the disease but are not able to provide an accurate prediction at a finer resolution. Factors, such as geographic location information, matter a lot in the process of virus transmission in the real world, which cannot be extracted from the statistical level data.

Another type of research utilizes individual level data to simulate the spread of infectious disease. With the popularity of personal devices, there are more and more research using individual level data such as mobile phone data \cite{bengtsson2015using,wesolowski2012quantifying,wesolowski2015impact} to analyze the impact of epidemic. Individual level data encodes more information than statistical level data, therefore these approaches generally represent the transmission better. This refined transmission model is better for analyzing the effectiveness of travel restrictions policies. Zhou et al. \cite{zhou2020effects} used the mobile phone sightings data provided by one of the mobile phone service providers in Shenzhen city to develop a modified SEIR model to analyze the effects of human mobility restrictions on the spread of COVID-19 in Shenzhen. But their model does not focus on the mapping of  simulation results of small world to real world.

Using individual level data, Small World Model method converts the result of small world into real world, analyzing the  effectiveness of different mobility restriction policies.

\section{Problem Description}

As for the simulation, this paper divides the whole city into cells with same area sizes. In every cell, the epidemic simulation is processed by running the SEIR model over time.

To map the simulation result of small world to the state of real world, Small World Model try to get the time scaling factor and number scaling factor, which indicates the ratio of disease spread timespan and the ratio of number of infected people between small world and real world. 
These two scaling factors link the spread of epidemics between  small world and real world.

The premise of the mapping from small world to real world is that the small dataset needs to meet the following two requirements:
\begin{itemize}
\item Data used in small world model are uniformly sampled from the real world population. In the simulation process, it can be considered that the mobility trajectory distribution in small world can represent the overall mobility trajectory distribution of real world.
\item The difference between real world and small world is population size, which affects population-related simulation parameters such as the total number of people per cell, the number of people per cell, the number of people in each time frame, and the frequency of contacts. Other parameters that are not related to the number of people are the same in real world and small world.
\end{itemize}



\section{Methodology}
\subsection{Epidemic transmission model}
SIR model, proposed by Kermack and McKendrick in 1927\cite{kermack1927contribution}, is one of the most famous epidemic transmission models. SIR model is a simple compartmental model. Many improved models such as SIS, SIRD and SEIR, are derivatives of the basic SIR. For many infectious diseases, there is a period of incubation during which individuals have been infected, but they cannot transmit the virus to other people. 
SEIR model takes the incubation period into account.
SEIR model consists of four compartments: susceptible (S), exposed (E), infectious (I), and recovered (R). Every susceptible individual (compartment S) becomes exposed (compartment E) at the moment of infection. The exposed individual will becomes infectious (compartment I) after the incubation period and then recovers (compartment R), with permanent or temporary acquired immunity.  

The SEIR model can be established as follows
\begin{equation}
\left\{
\begin{array}{rl}
\frac{dS}{dt} & = -\beta SI \vspace{1ex} \\ 
\frac{dE}{dt} & = \beta SI - \omega E \vspace{1ex} \\
\frac{dI}{dt} & = \omega E - \gamma I \vspace{1ex}\\
\frac{dR}{dt} & = \gamma I \vspace{1ex} \label{eq:seir}
\end{array} \right.
\end{equation}
where $\beta$ is the coefficient of infection rate, $\omega = \frac{1}{T_e}$, $\omega$ is the coefficient of migration rate of latency, $T_e$ is the average latency, $\gamma = \frac{1}{T_r}$, $\gamma$ denotes the coefficient of migration rate, and $T_r$ is the average recovery time.


\subsection{Time scaling factor and number scaling factor}
From the equation \ref{eq:seir}, the infection probability of susceptible populations is closely related to the number of contacts between peoples. In small world models, population density in each cell is lower than that of real world. Because the smaller population in the small world, the epidemic in small world will be at a slower rate than that of real world. The slowness is mainly reflected in the time and the number of people. That is why we need the time scaling factor and the number scaling factor.

To simplify the subsequent derivation, we use a new variable $IDI$ to denote the average crowd flow of each cell in each time frame in the model, as shown in the equation \ref{eq1}.
\begin{equation}
IDI=\frac{\sum_{i\in M}c_i}{N_{cell}}=\frac{M*Avg_c}{N_{cell}}
\label{eq1}
\end{equation}
where $N_{cell}$ is the total number of cells in the experimental area, $M$ is the population number of the small dataset, $c_i$ is the number of different cells that $user_i$ passes through in a specified time period. 
$\sum_{i\in M}c_i$ can be simplified to $M*Avg_c$, the product of the population number of the small world $M$ and the average number of visited cell per person $Avg_c$ in real world.


Based on the equation \ref{eq1}, the probability $\rho_1$ of a particular person go to a particular cell within the specific time period is given by the equation \ref{eq2}.

\begin{equation}
\rho_1=\frac{IDI}{M}=\frac{Avg_c}{N_{cell}} 
\label{eq2}
\end{equation}

Obviously, the probability $\rho_2$ of any two people visit a specified cell in the same time period is shown in the equation \ref{eq3}. 

\begin{equation}
\rho_2=\rho_1^2=\frac{Avg_c^2}{N_{cell}^2}
\label{eq3}
\end{equation}

Because the population in the small world model is uniformly sampled from the real world, the two probabilities $\rho_1$ and $\rho_2$ do not change as the dataset size changes. They can be considered as special constants.

However, the population $M$ is different in different dataset, in other words, in different world models. Considering the contact between people, each model has a total of $M(M-1)\approx M^2$ pairs of people. Therefore, considering the total number of people in the model, the average pairs in a specific cell in each time window is shown in the equation \ref{eq4}, which is converted to the $IDI$ related expressions. 

\begin{equation}
Conn\approx \rho_2M^2=IDI^2
\label{eq4}
\end{equation}

The equation \ref{eq4} only considers people visiting the same cell in the same time window, that is, only contacts but not close contacts. People who go to the same cell do not necessarily have close contact. 
The probability of close contact is positively related to the population density in the cell. That is the close contact occurrence rate $\rho_3 \propto IDI$. 

Therefore, in the time dimension, the spread speed of the epidemic $T$ is positively correlated with $IDI^3$:
\begin{equation}
T\propto IDI^3
\label{eq5}
\end{equation}

Since the positive correlation in equation \ref{eq5} does not contain elements related to the total population $M$ or other elements that cause differences between different worlds,
equation \ref{eq5} can be applied to different models. 
Further more, to compare the state under different policies, the specific comparison equation is shown in equation \ref{eq6}.
\begin{equation}
\frac{T_a}{T_b} = \frac{IDI_a}{IDI_b}
\label{eq6}
\end{equation}
where $T_a$ and $T_b$ correspond to the epidemic spread speed in policy $a$ and policy $b$, $IDI_a$ and $IDI_b$ are the average crowd flow per cell in two different models.




In addition to the impact of time scales, there are differences in the number of infected people between the small world model and the real world. In order to simplify the calculation, based on the SEIR model in the equation \ref{eq:seir}, there is a Markov state transition process between the number of infections at the previous moment and the number of infections at the new moment. We denote the transition process with a Markov state transition matrix $MK(t)$ here.
Since the infection rate used by the small world model is calculated based on the state of the real world, the Markov state transition matrix is consistent in the simulation of different small world model and the real world. Therefore, the change in the number of infected people can be simplified as shown in the equation \ref{eq7}.
\begin{equation}
Sn_t=Sn_{t-1}*MK(t)
\label{eq7}
\end{equation} 
where $Sn_{t-1}$ is the the total number of infected people at time $t-1$ in the small world model.

On the premise that the time scale is consistent, real world  should follow the infection rate similar to that in the small world model. Let the number of contacts in the real world be $r$ times the number of contacts in the small world model. As shown in the equation \ref{eq8}, the number of infected people in the real world $Rn_t$ can be directly converted from the number of infected people in the small world.

\begin{equation}
Rn_t=Sn_{t-1}*MK(t)*(1+r) = Sn_{t}*(1+r)
\label{eq8}
\end{equation}

However, the epidemic statistics of the real world are usually updated once a day, while the number of infected people can be updated once in a time slice in the simulation of the small world. In other words, the infected number in small world model $Sn_{t}$ is not equal to that in real world $Rn_t$ in the time dimension. To solve this problem, we use the compound interest formula \cite{sonin2020some} in economics to calculate the transfer relationship between $Rn_t$ and $Sn_t$ in a more detailed time slice. We can get the equation \ref{eq9}.
\begin{equation}
Rn_t=Sn_{t-1}*MK(t)*(1+\frac{r}{n})^n \approx Sn_{t-1}*MK(t)*e^r
\label{eq9}
\end{equation}
where $n$ is the multiplier between time frames of two worlds. For example, if the real world update time is 1 day and the small world model update time is 1 minute, here the multiplier $n=1440$.
The ratio of close contact persons $r$ in the equation \ref{eq9} is proportional to $\frac{IDI_a}{IDI_r}$. That is, $r\propto \frac{IDI_a}{IDI_r}$, where $IDI_a$ is the average crowd flow per cell in the small world model adopting travel restriction policy $a$, $IDI_r$ is the average crowd flow per cell in real world. 

From the equation \ref{eq9}, we can get the equation \ref{eq10}.
\begin{equation}
F = \frac{Rn_t}{Sn_t} = e^r
\label{eq10}
\end{equation}
where $F$ is the number scaling factor between real world and small world.

Based on the equation \ref{eq10}, to compare infected number in different travel restriction policy, the unified way of the number of infected people is shown in the equation \ref{eq11}. 
\begin{equation}
\frac{F_a}{F_b}=e^k(IDI_a-IDI_b)
\label{eq11}
\end{equation}
where $k$ is a constant, $IDI_a$ and $IDI_b$ are the $IDI$ in the travel restriction policy $a$ and policy $b$, $F_a$ and $F_b$ are the number scaling factors corresponding to $IDI_a$ and $IDI_b$.

Based on the equation \ref{eq6} and equation \ref{eq11}, the comparison of the effectiveness of the small world model under different travel restriction policies is defined.
\section{Conclusion}
Based on the small dataset, Small World Model is proposed to convert the simulation of small world model into the state of real world. Small World Model computes the number scaling factor and the time scaling factor to connect the simulation of small world and the state of real world. 
At the same time, using individual mobility data to simulate disease transmission under different policies, Small World Model can analyze the effect of different mobility restriction policies.

Small World Model is a general guidance that can represent the big world from the analysis of the small world, and it is not limited to simulation of the epidemic transmission. In the future, the team will expand Small World Model, through the analysis of the small world, so that it can reflect the mobility pattern of the real world.

\bibliographystyle{IEEEtran}
\bibliography{ref}

\begin{thebibliography}{10}
\providecommand{\url}[1]{#1}
\csname url@samestyle\endcsname
\providecommand{\newblock}{\relax}
\providecommand{\bibinfo}[2]{#2}
\providecommand{\BIBentrySTDinterwordspacing}{\spaceskip=0pt\relax}
\providecommand{\BIBentryALTinterwordstretchfactor}{4}
\providecommand{\BIBentryALTinterwordspacing}{\spaceskip=\fontdimen2\font plus
\BIBentryALTinterwordstretchfactor\fontdimen3\font minus
  \fontdimen4\font\relax}
\providecommand{\BIBforeignlanguage}[2]{{%
\expandafter\ifx\csname l@#1\endcsname\relax
\typeout{** WARNING: IEEEtran.bst: No hyphenation pattern has been}%
\typeout{** loaded for the language `#1'. Using the pattern for}%
\typeout{** the default language instead.}%
\else
\language=\csname l@#1\endcsname
\fi
#2}}
\providecommand{\BIBdecl}{\relax}
\BIBdecl

\bibitem{kraemer2019utilizing}
M.~Kraemer, N.~Golding, D.~Bisanzio, S.~Bhatt, D.~Pigott, S.~Ray, O.~Brady,
  J.~Brownstein, N.~Faria, D.~Cummings \emph{et~al.}, ``Utilizing general human
  movement models to predict the spread of emerging infectious diseases in
  resource poor settings,'' \emph{Scientific reports}, vol.~9, no.~1, pp.
  1--11, 2019.

\bibitem{charaudeau2014commuter}
S.~Charaudeau, K.~Pakdaman, and P.-Y. Bo{\"e}lle, ``Commuter mobility and the
  spread of infectious diseases: application to influenza in france,''
  \emph{PloS one}, vol.~9, no.~1, p. e83002, 2014.

\bibitem{mari2012role}
L.~Mari, E.~Bertuzzo, L.~Righetto, R.~Casagrandi, M.~Gatto,
  I.~Rodriguez-Iturbe, and A.~Rinaldo, ``On the role of human mobility in the
  spread of cholera epidemics: towards an epidemiological movement ecology,''
  \emph{Ecohydrology}, vol.~5, no.~5, pp. 531--540, 2012.

\bibitem{chinazzi2020effect}
M.~Chinazzi, J.~T. Davis, M.~Ajelli, C.~Gioannini, M.~Litvinova, S.~Merler,
  A.~P. y~Piontti, K.~Mu, L.~Rossi, K.~Sun \emph{et~al.}, ``The effect of
  travel restrictions on the spread of the 2019 novel coronavirus (covid-19)
  outbreak,'' \emph{Science}, vol. 368, no. 6489, pp. 395--400, 2020.

\bibitem{kucharski2020early}
A.~J. Kucharski, T.~W. Russell, C.~Diamond, Y.~Liu, J.~Edmunds, S.~Funk, R.~M.
  Eggo, F.~Sun, M.~Jit, J.~D. Munday \emph{et~al.}, ``Early dynamics of
  transmission and control of covid-19: a mathematical modelling study,''
  \emph{The lancet infectious diseases}, vol.~20, no.~5, pp. 553--558, 2020.

\bibitem{fang2020transmission}
Y.~Fang, Y.~Nie, and M.~Penny, ``Transmission dynamics of the covid-19 outbreak
  and effectiveness of government interventions: A data-driven analysis,''
  \emph{Journal of medical virology}, vol.~92, no.~6, pp. 645--659, 2020.

\bibitem{liu2020modelling}
M.~Liu, J.~Ning, Y.~Du, J.~Cao, D.~Zhang, J.~Wang, and M.~Chen, ``Modelling the
  evolution trajectory of covid-19 in wuhan, china: experience and
  suggestions,'' \emph{Public health}, vol. 183, pp. 76--80, 2020.

\bibitem{wells2020impact}
C.~R. Wells, P.~Sah, S.~M. Moghadas, A.~Pandey, A.~Shoukat, Y.~Wang, Z.~Wang,
  L.~A. Meyers, B.~H. Singer, and A.~P. Galvani, ``Impact of international
  travel and border control measures on the global spread of the novel 2019
  coronavirus outbreak,'' \emph{Proceedings of the National Academy of
  Sciences}, vol. 117, no.~13, pp. 7504--7509, 2020.

\bibitem{bengtsson2015using}
L.~Bengtsson, J.~Gaudart, X.~Lu, S.~Moore, E.~Wetter, K.~Sallah, S.~Rebaudet,
  and R.~Piarroux, ``Using mobile phone data to predict the spatial spread of
  cholera,'' \emph{Scientific reports}, vol.~5, no.~1, pp. 1--5, 2015.

\bibitem{wesolowski2012quantifying}
A.~Wesolowski, N.~Eagle, A.~J. Tatem, D.~L. Smith, A.~M. Noor, R.~W. Snow, and
  C.~O. Buckee, ``Quantifying the impact of human mobility on malaria,''
  \emph{Science}, vol. 338, no. 6104, pp. 267--270, 2012.

\bibitem{wesolowski2015impact}
A.~Wesolowski, T.~Qureshi, M.~F. Boni, P.~R. Sunds{\o}y, M.~A. Johansson, S.~B.
  Rasheed, K.~Eng{\o}-Monsen, and C.~O. Buckee, ``Impact of human mobility on
  the emergence of dengue epidemics in pakistan,'' \emph{Proceedings of the
  National Academy of Sciences}, vol. 112, no.~38, pp. 11\,887--11\,892, 2015.

\bibitem{zhou2020effects}
Y.~Zhou, R.~Xu, D.~Hu, Y.~Yue, Q.~Li, and J.~Xia, ``Effects of human mobility
  restrictions on the spread of covid-19 in shenzhen, china: a modelling study
  using mobile phone data,'' \emph{The Lancet Digital Health}, vol.~2, no.~8,
  pp. e417--e424, 2020.

\bibitem{kermack1927contribution}
W.~O. Kermack and A.~G. McKendrick, ``A contribution to the mathematical theory
  of epidemics,'' \emph{Proceedings of the royal society of london. Series A,
  Containing papers of a mathematical and physical character}, vol. 115, no.
  772, pp. 700--721, 1927.

\bibitem{sonin2020some}
I.~M. Sonin and M.~Whitmeyer, ``Some nontrivial properties of a formula for
  compound interest,'' \emph{Finance Research Letters}, vol.~33, p. 101217,
  2020.

\end{thebibliography}

\end{document}